%% file: manuscript.tex
\documentclass[fleqn,usenatbib]{mnras}

\usepackage{newtxtext,newtxmath}
\usepackage[T1]{fontenc}
\usepackage{ae,aecompl}

\input{packageIncludes}

\newcommand{\msun}{\ensuremath{\mathrm{M_{\odot}}}}

\input{figureMacros}
\DeclareSIUnit[number-unit-product = \;]
\solarmass{\msun}

\DeclareSIUnit[number-unit-product = \;]
\dyn{dyn}

\DeclareSIUnit[number-unit-product = \;]
\ergy{erg}

\hypersetup{
    colorlinks,
    citecolor=blue,
    filecolor=black,
    linkcolor=blue,
    urlcolor=black
}

\title[Detonability of white dwarf plasma]
      {Detonability of white dwarf plasma: turbulence models at low densities}

\author[D. Fenn and T. Plewa]
{D. Fenn$^{1}$\thanks{E-mail:df11c@my.fsu.edu (DF); tplewa@fsu.edu (TP)} and
T. Plewa$^{1\ast}$\thanks{corresponding author}\\
$^{1}$Department of Scientific Computing, Florida State University, Tallahassee, FL 32306, U.S.A.}

\begin{document}

\date{Accepted 2017 February 27. Received 2017 February 23; in original form 2016 December 21}

\pagerange{{1361}--{1372}} \pubyear{2017} \volume{468}

\label{firstpage}
\maketitle

\begin{abstract}
We study the conditions required to produce self-sustained detonations
in turbulent, carbon-oxygen degenerate plasma at low densities.
We perform a series of three-dimensional hydrodynamic simulations of
turbulence driven with various degrees of compressibility. The average
conditions in the simulations are representative of models of merging
binary white dwarfs.
We find that material with very short ignition times is abundant in
the case that turbulence is driven compressively. This material forms
contiguous structures that persist over many ignition times, and that
we identify as prospective detonation kernels. Detailed analysis of
prospective kernels reveals that these objects are centrally-condensed
and their shape is characterized by low curvature, supportive of
self-sustained detonations. The key characteristic of the newly
proposed detonation mechanism is thus high degree of compressibility
of turbulent drive.
The simulated detonation kernels have sizes notably smaller than the
spatial resolution of any white dwarf merger simulation performed to
date. The resolution required to resolve kernels is \SI{0.1}{\km}. Our
results indicate a high probability of detonations in such
well-resolved simulations of carbon-oxygen white dwarf mergers. These
simulations will likely produce detonations in systems of lower total
mass, thus broadening the population of white dwarf binaries capable
of producing Type Ia supernovae.  Consequently, we expect a downward
revision of the lower limit of the total merger mass that is capable
of producing a prompt detonation.
We review application of the new detonation mechanism to various
explosion scenarios of single, Chandrasekhar-mass white dwarfs.

\end{abstract}


%%%%%%%%%%%%%%%%%%%%%%%%%%%%%%%%%%%%%%%%%%%%%%%%%%

%%%%%%%%%%%%%%%%% BODY OF PAPER %%%%%%%%%%%%%%%%%%
%
%
%
\section{Introduction}\label{s:intro}
The origins of Type Ia supernovae (SNe Ia), important for their use as
cosmological distance indicators, remains an unsolved problem. Decades
of research indicate that these events are produced by thermonuclear
runaways in the degenerate material of white dwarfs
\citep{Hoyle_Fowler}; however, the mechanism of these explosions and
the nature of the progenitor systems remain unknown. These missing
details are vital components of the initial conditions of the SN Ia
explosion problem, and therefore are essential to understanding these
events.

A requirement of any successful SN Ia model is that a sufficient
portion of the white dwarf material must be brought under conditions
of densities and temperatures sufficiently high to enable explosive
burning. There are many theories as to how this may happen; we briefly
review here the most important in the context of the present work.
            
The single degenerate (SD) SN Ia explosion scenario concerns a binary
system consisting of a white dwarf accompanied by a non-degenerate
star. This model posits that the white dwarf slowly gains mass as it
accretes material from its companion. Eventually, a thermonuclear
runaway occurs as the white dwarf nears the Chandrasekhar mass. It is
theorized that mild ignition of carbon initiates an explosion that
first proceeds as a deflagration that ultimately consumes the star
\citep{nomoto+76, nomoto+84}. Because pure deflagration models have
difficulty in explaining observations, it has been proposed that a
deflagration-to-detonation transition, or DDT \citep[see][for review]
{oran+07}, occurs upon the flame reaching the low-density outer layers
of the white dwarf \citep[see][and references
  therein]{gamezo+05}. These delayed-detonation models show close
agreement with observations for a range of SN Ia types
\mbox{\citep{blondin+13}}, although the postulated mechanism behind DDT
remains unexplained (see below).

An additional source of SN Ia events may also be binary white dwarf
systems. In this case, a supernova explosion occurs when two white
dwarfs merge as the binary system loses orbital angular momentum due
to emission of gravitational waves. This formation channel
is known as the double degenerate (DD) scenario. Here, the ignition of
the stellar plasma takes place either early on during the violent
phase of the merger \citep{pakmor+10, guillochon+10}, in the hot torus
soon after the merger process is complete \citep{mochkovitch+89}, or
only after the merged object cools sufficiently and compresses in the
core \citep[][see, e.g.,]{raskin+2014}.

Aside from the nature of the progenitor system, there remains
considerable uncertainty regarding the physical mechanism underlying
ignition in SN Ia. In this work, we are primarily concerned with the
mechanism of self-sustained detonations, which may be either directly
initiated or formed in the process of DDT. In the former case, a shock
compresses and heats thermonuclear fuel, causing it to explosively
burn, with the energy spent on supporting the expanding detonation
wave \citep[see, e.g.,][]{dursi+06}. In the merger scenario, the shock
that triggers the detonation may originate in the process of
amplification of acoustic perturbations in the hot, turbulent boundary
layer that forms around the primary component.

In the case that no shocks are present, detonations can be initiated
via the Zel'dovich mechanism \citep{zeldovich+70}, in which an
acoustic perturbation is gradually strengthened thanks to an existing
reactivity gradient. This mechanism has been considered as a
particular source of DDT in the single-degenerate scenario
\citep[][and references therein]{khokhlov+97,woosley+2011}. In the
case that the medium is inhomogeneous, the detonation may emerge as
the result of a sequence of discrete ignition events, as described in
the SWACER mechanism \citep{khokhlov91b,oran+07}. More recently, based
on the results of direct numerical simulations of turbulent chemical
combustion, it has been suggest that a special preconditioning of the
medium may not be required for DDT to occur \citep{poludnenko+11}. In
this model, DDT is due to an inherent instability of flames to
turbulence, provided turbulence is sufficiently strong.

Regardless of the scenario of detonation initiation, specific
conditions must be satisfied in order for the detonation to propagate
in a self-sustained way. For example, the boundary of the detonating
region must have a sufficiently small curvature \citep{sharpe+01}. (In
the context of astrophysical applications, the maximum curvature
condition is frequently expressed in terms of a minimum critical size
of the detonating region \citep{dursi+06,dunkley+2013}.)
Additionally, ignition times of the mixture must be shorter than the
characteristic time over which the fluid mixes or expands. Otherwise,
ignition kernels will be either structurally destroyed or the plasma
burning will cease due to adiabatic cooling. Ultimately, an energetic
Type Ia supernova event requires a detonation to burn oxygen-rich
material \citep{seitenzahl+2009, fenn+2016}.

A recent comprehensive survey of a parameter space of white dwarf
mergers by \cite{dan+14} indicated that successful supernova
explosions during the initial, violent merger phase might be produced
only in systems with a total mass in excess of about 2.1 \msun. Their
conclusion appears supported by a number of merger simulations
obtained by other groups \citep[see, e.g.,][]{kashyap+2015,
  fenn+2016}. If true, this would presumably eliminate the possibility
of observing substantial asymmetries in all Type Ia supernovae, as
such systems could only be produced in the course of prompt
explosions. Indeed, SNe Ia polarization observations indicate that
these objects are nearly spherically symmetric. If indeed massive
merger systems explode promptly, as indicated by computer simulations,
it is then conceivable that the degree of observed asymmetry should
correlate with the mass of the exploding merger object. That is, very
luminous Type Ias produced as a result of prompt detonations in
massive mergers should be characterized by a significant amount of
polarization. Therefore, it is important to understand the conditions
under which prompt detonations occur in merging white dwarf
binaries. This in turn would allow to constrain the (minimal) total
mass of the supernova progenitor system.

Prompt detonations can be produced during the merger process in either
the boundary layer formed around the primary component, or in the core
of the primary. In the latter case, the mechanism responsible for
ignition is tidal compression \citep{fenn+2016}. In the case of
boundary-layer ignition, the suspected mechanism has not always been
clearly identified \citep{moll+2014, kashyap+2015}, and resolution
achievable in merger simulations might be inadequate to correctly
capture the important physics of the problem \citep{fenn+2016,
  yungelson+17}. Our somewhat better resolved simulations of mergers
\citep{fenn+2016} provided strong evidence that the combined effects
of turbulence and thermonuclear self-heating lead to detonations in
helium-rich material. In this case, the observed evolution toward
detonation resembled a process of gradual strengthening of the
spontaneous wave \citep{khokhlov91}, but detailed analysis of the
process was outside the scope of our study. Ultimately, however, the
helium detonation proved insufficient to produce a supernova explosion
(detonate carbon) during the violent phase of the merger.

Our aim is to identify the conditions required to trigger detonations
in turbulent, carbon-rich degenerate plasma at low densities. This
problem has numerous applications in the context of Type Ia
supernovae. First, as discussed above, it will help to answer the
question of whether prompt detonations can be produced in mergers of
carbon-rich white dwarfs (CO/CO). Second, a similar question is
relevant in the context of sub-Chandrasekhar systems with surface
helium detonations (He/CO). Third, carbon detonations are expected to
be produced in the course of the DDT mechanism in Chandrasekhar-mass
white dwarfs \citep{khokhlov91}. Also, we want to provide evidence for
limitations of current integrated computer models of white dwarf
mergers and establish a reference point for future simulations that
require capturing the effects of turbulent thermonuclear burning.

We study this problem by modeling the hydrodynamic evolution of a
turbulent region filled with a carbon/oxygen mixture at a density,
temperature, and Mach number characteristic of conditions found in the
$0.8 + \SI{1.2}{\solarmass}$ (CO0812) model by
\citet{fenn+2016}. Thus, the plasma is mildly degenerate and, on
average, thermonuclear burning is inefficient. However, as the region
becomes turbulent, one expects fluctuations to develop around the
average hydrodynamic state. The model outputs of interest in this case
and in the context of detonations are the distributions of ignition
times, their temporal evolution, and the masses and structures of
regions where ignition times are sufficiently short for detonations to
occur. We identify the latter regions as prospective detonation
kernels. These kernels must satisfy certain mass and morphological
criteria in order to produce self-sustained detonations. We probe the
ignition conditions for different amounts of compressibility in
turbulent driving. It is expected that, for a given driving energy,
the density and temperature fluctuations, and thus, fluctuations in
ignition times, will grow in amplitude as the driving becomes
progressively more compressive. We asses the probability of successful
detonations based on characteristics of the identified prospective
detonation kernels.
\section{Models and methods}
Although we are concerned with a problem that is more broadly
relevant, in this study we consider conditions representative of the
boundary layer that forms around the primary component in the process
of merging of two white dwarfs. This simplifies our computational
setup as we can use one simulation code, a single computational
domain, and there are only a few problem control parameters. We begin
description of our model by presenting details of computational model
and then provide a description of a set of tools used in analysing
simulation results.
\subsection{Computational method}
We perform a set of three-dimensional simulations of driven turbulence
using \textsc{Proteus}, a fork of the \textsc{FLASH} code
\mbox{\citep{Fryxell+00}}. The time-dependent Euler equations are
solved using the Piecewise Parabolic Method
\mbox{\citep{colella+84}}. We seed the simulations with passive
Lagrangian tracer particles in order to enable Lagrangian flow
diagnostics. We use the Helmholtz equation of state
\mbox{\citep{timmes+2000}}.
\paragraph{Computational domain and boundary conditions}
The simulation domain is a cube with sides of length \SI{3.2e6}{\cm}
in Cartesian geometry, uniformly refined with 512 mesh zones per
dimension. Given the relatively small size of the considered region
compared to the dimensions of the boundary layer, we impose periodic
boundary conditions.
\paragraph{Initial conditions}
We use the CO0812 model of \citet{fenn+2016} to determine initial
conditions for the simulations presented here. Our choice follows the
somewhat unexpected simulation result that no detonation was observed
in the boundary layer of this model. At the initial time, the
simulation box is filled with plasma composed of half carbon and half
oxygen at a constant density of \SI{4e6}{\g\per\cubic\cm} and
temperature \SI{1e9}{\K}. It should be noted that we do not account
for compositional changes that would have occurred as the material was
heated to the temperature used here (for the conditions considered
here, less than 10 per cent of the carbon is burned to heat the plasma
to the adopted initial temperature).
\paragraph{Spectral driving of turbulence}
The turbulence in our simulations is driven using the spectral forcing
method first described by \mbox{\citet{eswaran+88}}, in which kinetic
energy is added to the system in large scale (low wavenumber) patterns
for wavenumbers ranging from $k=1$ to $k=4$, with a decay time
comparable to the sound-crossing time of the domain
(\SI{0.01}{\s}). The details of the driving procedure generalized to
enable an arbitrary degree of compressibility are given in
\citet{federrath+2010}. The amount of the injected turbulent
energy was tuned in such a way that the domain-averaged Mach number
was approximately equal to 0.5, matching that of the CO0812 model
of \citet{fenn+2016}.

The driving energy required to achieve this average Mach number is
dependent on the degree of compressibility of the turbulent
forcing. We define the quantity $\zeta_c = 1-\zeta$, where $\zeta$ is
the forcing parameter given by \citet{federrath+2010}. In
this notation, $\zeta_c=0$ represents the degree of compressibility of
the forcing, so that $\zeta_c=0$ corresponds to purely solenoidal
forcing, and $\zeta_c=1$ to purely compressive forcing. We label each
of our models as $C_{\zeta_c}$; thus, the model $C_0$ uses only
solenoidal forcing, and $C_1$ is driven compressively. The models
used, as well as the corresponding values for the turbulent driving
energy, are summarized in \mbox{Table~\ref{t:parameterSpace}}.
\ctable[
cap = turbulence parameters,
caption = {Model designations and parameters of the turbulent driving.} ,
label = t:parameterSpace,
mincapwidth = 8cm,
star
]{l|lllll}
{
}
{\FL
model & $C_0$ & $C_{0.5}$ & $C_{0.75}$ & $C_{0.875}$ & $C_{1}$ \NN
$\zeta$ & 0 & 0.5 & 0.75 & 0.875 & 1                    \NN
$E_{\mathrm{d}}$ [erg] & \num{4e15} & \num{5e15} & \num{2e16} & \num{8e16} & \num{1e18}  \LL
}
We study the evolution of the model systems for an extent of time
comparable to the dynamical (free-fall) timescale of the average
plasma density, which is generally less than 1 second. Our previous
simulations showed that the average change in boundary layer
temperatures over these timescales was negligible. However, a small
amount of the energy injected in the process of driving turbulence is
thermalized, causing the average simulation conditions, especially
temperature, to change. In order to prevent this from happening, we
remove the excess internal energy from the system after each
simulation step in such a way that the total system internal energy
remains constant throughout the simulation. This procedure also
ensures that the average temperature of the system remains constant
over time.
\subsection{Analysis methods}
\subsubsection{Characterization of turbulence}
In order to verify the correct performance of our driven turbulence
model, we performed a set of trial simulations at resolutions of
$128^3$, $256^3$, and $512^3$ using solenoidal forcing, and an
additional simulation with resolution $512^3$ using purely compressive
forcing. We determined the integral length and time scales in this set
of simulations by integrating the respective spatial and temporal
autocorrelation functions \citep{davidson2004}. In the case of
solenoidal stirring, the integral time and length scales were found to
be $\approx \SI{3e-3}{\s}$ and $\approx
  \SI{2e5}{\cm}$, respectively. We confirmed these estimates
independently by integrating the kinetic energy spectra
\mbox{\citep[see equation 13 in][]{eswaran+88}}, and obtained
agreement to within a factor of about 2.

We additionally performed independent estimates of the integral time
scale by calculating the quantity $l/u$, where $l$ is the integral
length scale, and $u$ is the characteristic large-scale velocity.
Here the values of $u$ were determined using two different methods. In
the first method, we used the volume RMS velocity; in the second
method, we used a Fourier transform analysis to remove small- and
medium-scale velocity components from the velocity spectrum. In all
cases, agreement with estimates obtained using the autocorrelation
functions was within a factor of 2.

As we are chiefly interested in steady-state properties of the
turbulence, results presented in this study exclude the initial
transient phase (during which the turbulence is not fully
developed). Observations of both the average Mach number and system
enstrophy indicated that the system is in a quasi-steady state after
about 25 integral time scales. The simulations were run for about 15
integral time scales in quasi-steady state, in accordance with the
condition that the simulation should encompass many large-eddy
turnover times \mbox{\citep{eswaran+88}}.

\begin{figure*}
\adjustTwoPanels\ignorespaces
\begin{center}
    \begin{tabular}{cc}
                \includegraphics{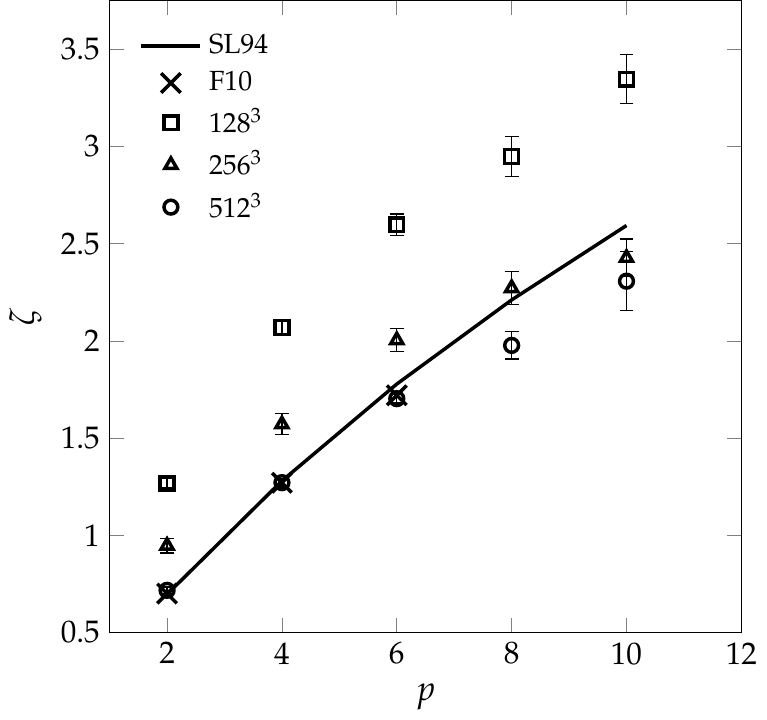}
                &
                \includegraphics{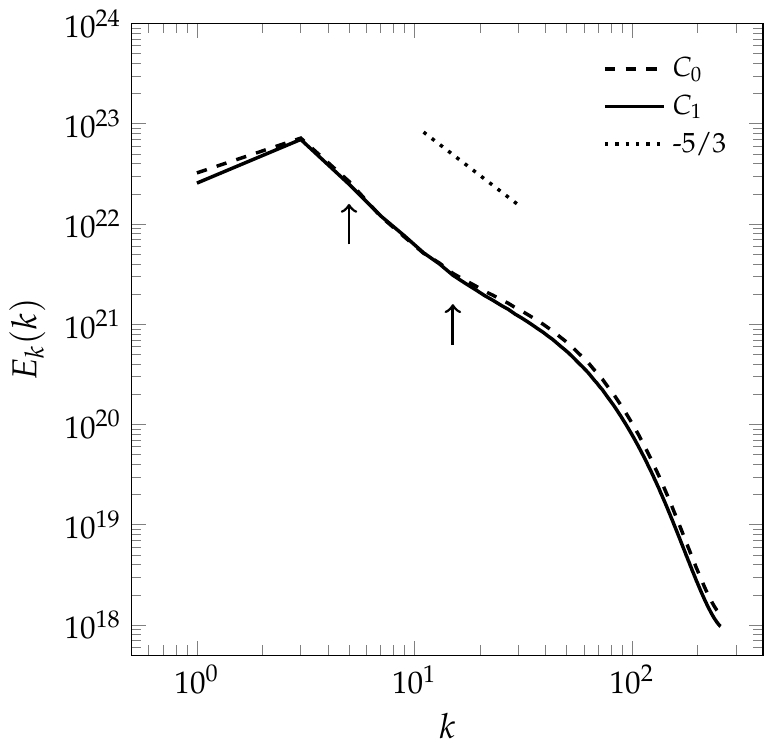}
    \end{tabular}
    \caption
    { Left panel: structure function scaling exponent,
        $\zeta$, as a function of structure function order,
        $p$, for mesh resolutions of $128^3$, $256^3$, and
      $512^3$. The error bars indicate 95\% confidence intervals for
      the exponent fits. The model by \citet{she+94} is
      provided for reference (black line). Additionally, the
      results from \citet{federrath+2010} are included for
      comparison. Our results show good agreement with
      \citet{federrath+2010}, though both deviate from the
      \citet{she+94} prediction for high values of $p$. Right panel:
      the turbulent kinetic energy spectrum, $E_k(k)$, for
      solenoidally (dashed line) and compressively (solid line) driven
      turbulence. The dotted line gives the Kolmogorov -5/3 slope for
      reference, and the arrows indicate the limits of the inertial
      range. The spectra for both driving modes are very similar, and
      the inertial range slopes agree well with those obtained by
      \citet{federrath+2010} (see text). \label{f:turbVerification} }
\end{center}
\end{figure*}

We summarize the results of our trial driven turbulence study by
presenting the structure function scaling exponents and the turbulent
kinetic energy spectra. The structure function scaling exponents are
shown in the left panel of Figure~\ref{f:turbVerification}
%
%
%
% Two panels:
% 1: She-Leveque fits for different resolutions
% 2: KE spectra for solenoidal and compressive stirring
for different model resolutions. The results indicate that the
exponents converge as the resolution increases, and the exponents from
our best-resolved $512^3$ model agree well with those obtained by
\citet{federrath+2010}. Also, the dependence of the scaling exponents
on the structure function order obtained from our simulations is
characterized by a shallower slope than predicted by the
\citet{she+94} model. This is consistent with other studies of
compressible turbulence \citep[see, e.g.,][]{fisher+2008}. As pointed
out by \citet{federrath+2010}, based on simulations obtained at still
higher resolutions, simulations with a resolution of $512^3$ can be
considered converged. This is because at this resolution numerical
errors are smaller than the scale of the natural variance inherent to
turbulent flows, and thus have a small effect on solution
accuracy. Subsequently, we adopted a resolution of $512^3$ as the
standard resolution in our simulations.

The time-averaged turbulent kinetic energy spectra obtained in
simulations with solenoidal forcing (shown with a dashed line) and
compressive forcing (shown with a solid line) are depicted in the
right panel of Figure~\ref{f:turbVerification}. The spectra display
the characteristics typical of driven turbulence, with the energy
injected on large scales ($1 \le k \le 4$ in our models) with the
energy cascading down to smaller scales. We calculated the inertial
range slopes for the region $5 \le k \le 15$, chosen to coincide with
the inertial range used by \citet{federrath+2010}. This gave
values of the slope in the inertial range of -1.90 for solenoidal
forcing, and -1.89 for compressive forcing. These indicate a steeper
dependence than the canonical value of -5/3 for Kolmogorov turbulence
(shown with a dotted line in the right panel of
Figure~\ref{f:turbVerification}). These exponents closely match the
results reported by \citet{federrath+2010}. We conclude that our trial
computer experiments produce results that are in good agreement with
the previously reported simulations.
\subsubsection{Analysis of ignition conditions} \label{ss:analysisIgCond}
In order for a parcel of degenerate matter to produce a self-sustained
detonation, it must be characterized by the following conditions: (1)
ignition times must be shorter than the local dynamical timescale
\citep{fowler+64,arnett96}, (2) the burning parcel must have a small
curvature on average \cite[sizable and of regular
  shape;][]{sharpe+01}, and finally, (3) the ignition times must also
be shorter than the local mixing timescale (we term this as
longevity). Conditions (1) and (2) have been extensively discussed in
the astrophysical literature \citep[see,
  e.g.,][]{dursi+06,seitenzahl+2009,dunkley+2013}, but we are not
aware of any study specifically addressing the longevity of
prospective detonation kernels, which would require sufficiently
realistic hydrodynamic simulations. The model employed here offers
such an opportunity, and having complete control of the flow
conditions in our model, we are well-positioned to address this aspect
in detail.

We developed a set of tools enabling analysis of the distribution of
ignition times from the perspective of the production of
self-sustained detonations. To this end, we identified and tracked
regions of interest characterized by relatively short ignition
times. At this point, we are not in a position to strictly define the
longest ignition time required to produce a self-sustained detonation,
as this time depends on several factors, as explained in the previous
paragraph. The detonation kernel analysis was performed only after the
system had reached a quasi-steady state, and was applied over
sufficiently long periods of time, as described below.
\paragraph{Clustering analysis} \label{p:clustering}
To identify prospective detonation kernels, defined as contiguous
regions with ignition times not exceeding a certain threshold value,
$\tau_{\mathrm{ign,pdk}}$, we masked cells on the hydro mesh that
satisfied this criterion. We then used a flood-filling algorithm
\citep[][and references therein]{levoy81} to composite (cluster) the
masked cells into contiguous regions. Because flood-filling algorithms
are typically implemented using a recursive procedure, special
consideration had to be taken to prevent exhausting memory made
available by the computer system to the application (stack overflow),
as the data sets are relatively large. We also note that our use of the
threshold ignition time in defining the spatial extent of the kernels
implies that we expect that ignition will occur anywhere inside this
region, and will result in a detonation (self-sustaining of
detonations requires additional characterization beside the ignition
time, as discussed previously). Naturally, it is expected that the
ignition times will vary inside the kernel, possibly attaining values
much shorter than the specified threshold. We omit such details of the
kernel structure from our considerations, as our focus is the
existence of the detonation kernels.

The clustering analysis was performed at a certain time, denoted
$t_{\mathrm{i,pdk}}$ here, from which the evolution of the prospective
kernels could be studied. In this work, we tracked their evolution
using both Eulerian and Lagrangian fluid flow representations, as
described below.
\paragraph{Longevity analysis} \label{p:longevity}
Our aim in this analysis step was to determine for how long the
originally-identified prospective kernels existed in the course of the
simulation. This information can be considered from the Eulerian as
well as Lagrangian points of view, and we used both approaches in the
present work.

In the first step of the Eulerian approach, every prospective kernel
was assigned a unique identifier. This identifier was then used as the
initial value of a new hydro variable (passive mass tracer), which was
advected with the flow. Thus, we created a set of mass tracers
uniquely representing material of prospective detonation kernels. As
the simulation progressed, at selected times for
$t>t_{\mathrm{i,pdk}}$, we repeated the clustering analysis and
identified new prospective kernels. We term these newly-identified
kernels. At this point of the analysis procedure, time-evolved
information about the distribution of the original prospective
detonation kernels as described by the mass tracers, as well as the
information about the newly-identified kernels is available.

Next, we must find which original kernels contribute material to the
newly-identified kernels. This step is necessary because, due to the
Eulerian character of our scheme, and numerical mixing, intermediate
values of mass tracers can be found in the material of the
newly-identified kernels. However, these new kernels are expected to
contain at least some material from the original kernels. Such
material can be relatively easily identified by looking for cells that
are filled with material from only one of the original kernels. Thus,
at this point we limit the origin of material of the newly-identified
kernels to only a subset of original kernels (in the limit, all the
original kernels may contribute to one or more of the new kernels).

In the final step of the Eulerian approach, we consider mixed cells
(those that contain various fractions of the original kernels). Here
the cells that contain at least 90 per cent of the specific original
cluster material are assigned to the surviving remnant of this
particular original cluster. This concludes the process of
(re)construction of the prospective kernels at the new time level.

In the Lagrangian approach, we seeded simulations with passive tracer
particles. Every particle was assigned an identifier corresponding to
its parent prospective detonation kernel. The sets of time slices of
particle distributions were used to calculate time-based velocity
autocorrelation functions, $\rho(t)$ \citep{tennekes+72}. The
autocorrelation functions were calculated individually for each
cluster. These functions provide information about the time coherence
of prospective kernels from the perspective of hydrodynamic mixing. We
chose not to account for the potentially changing ignition times (due
to expansion or contraction of material) in the Lagrangian analysis,
as it primarily serves to corroborate the results of the Eulerian
analysis.

Analysis of the prospective detonation kernels was continued either
for several $\tau_{\mathrm{ign,pdk}}$, as specified in the clustering
step described above, or until all prospective kernels had been
destroyed. We considered a prospective kernel destroyed when we could
no longer find on the mesh any cells that contained pure material from
the original kernel. The Lagrangian method was performed along with
the Eulerian analysis, and both were terminated at the same time.
\paragraph{Compactness analysis} \label{p:compactness}
An additional important consideration in the context of the ignition
of self-sustained detonations is the morphology of the prospective
detonation kernels. Stable detonations require continuous fronts of
sufficiently low curvature \citep{sharpe+01}. For simplicity, assuming
that a kernel is spherical in shape, the curvature of the kernel
surface (which is associated with the detonation front) inversely
scales with the kernel radius. This simplifying assumption is not
satisfied in general in hydrodynamic simulations of reactive flows,
and in particular in our models. For example, prospective kernels may
have the appearance of extended 2D-like structures, which cannot be
easily characterized as spheres or assigned a single curvature
value. We may measure by how much each kernel's morphology deviates
from spherical by calculating the kernel volume filling factor,
$f_\mathrm{pdk}$. The kernel's volume filling factor is defined as the
fraction of volume occupied by the kernel material inside a spherical
shell with a given inner and outer radius. The shell is
centred at the kernel's centre of mass. A filling factor of
less than 1 then indicates that the material is not spherically
arranged for that range of radii.

For each kernel, we calculated the dependence of the filling factor on
radius starting from the kernel's centre of mass and ranging to the
radius at which the filling factor was zero. The kernel compactness
can be understood in terms of the dependence of the volume filling
factor on the distance from the kernel's centre. A compact kernel
would be characterized by a relatively large filling factor at small
radii. Then the requirement of low curvature of the detonation front,
in our adopted simplified representation of spherical kernels, would
be satisfied by compact kernels with appropriately large radii. Only
such kernels may be considered as actual detonation kernels.
\paragraph{Fractal dimension estimation}
An additional measure of a kernel's morphology can be obtained by
estimating the fractal dimension of its surface. The fractal dimension
reflects on the object's morphological complexity, and more
specifically describes how a structure's detail changes with
scale. When a fractal analysis is applied to a kernel's surface, it
then characterizes how the amount of structure in the surface depends
on scale. Therefore, the greater the fractal dimension of a kernel's
surface, the greater the complexity of its shape. In the work
presented here, the kernel surface fractal dimension was estimated
using a box-counting method \citep{falconer+2003}.
\section{Results and Discussion}\label{s:resultsDiscussion}
As mentioned in Section~\ref{s:intro}, the successful formation of a
detonation in white dwarf material depends on several necessary
conditions relating to the minimum sizes of ignition kernels, their
morphology, and their corresponding ignition delay times. In this
section, we systematically examine the likelihood of explosive nuclear
burning for varying degrees of compressibility in the turbulent
driving. For the simulations using different stirring models
(characterized by a different degree of compressibility in the
turbulent driving, $\zeta_c$), we perform a clustering procedure which
identifies the material that has ignition times below a given
threshold and produces a set of separate, contiguous regions. If the
mass of the cluster exceeds the critical mass deemed necessary to
produce a self-sustained detonation \citep[see, e.g.,][]{dursi+06}, we
tentatively identify the cluster as an ignition kernel. This
identification is only tentative as such kernels must satisfy
additional criteria in order to produce self-sustained
detonations. First, the kernel material must not mix with the
surrounding medium for times shorter than the ignition time. That is,
the kernels must survive for long enough for the detonation to fully
develop. In addition, the ignition kernels should be compact. That is,
they must be approximately spherically symmetric in order for a
detonation to be uniformly powered by the burning zone.
\subsection{Mass distribution of ignition times} \label{s:massDistributions}
\begin{figure}
    \centering    
    \includegraphics{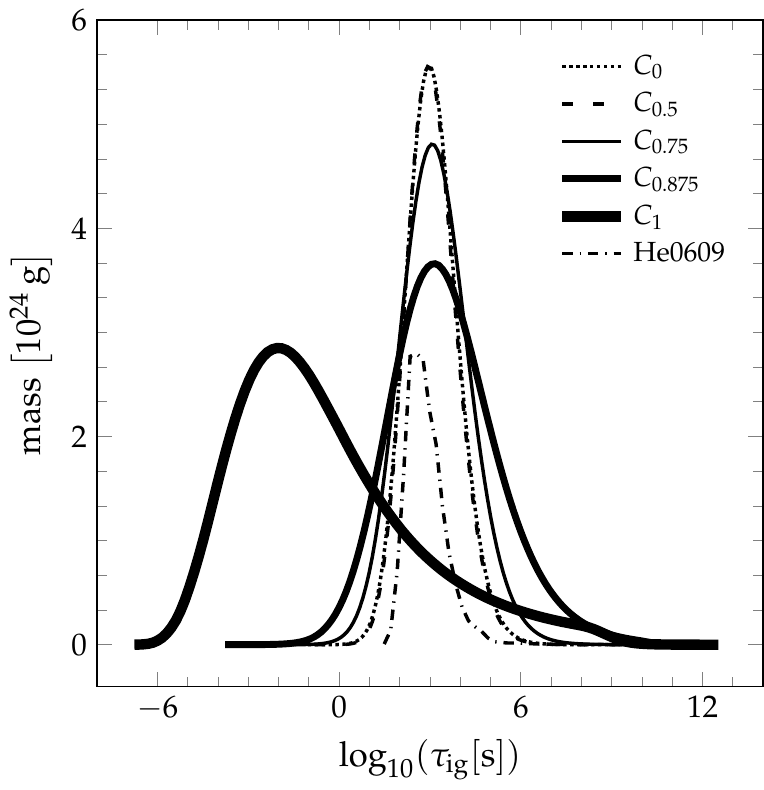}
    \caption
    { Mass distributions as a function of carbon ignition time for
      various values of $\zeta_c$. The distributions for the $C_0$ and
      $C_{0.5}$ models are visually indistinguishable. Increased
      compressibility in the drive causes the distributions to spread
      and develop more extended tails, indicating greater extremes in
      ignition times. Also shown is the mass distribution of helium
      ignition time calculated from a cube of boundary-layer data
      extracted from the He0609 model of \citet{fenn+2016}
      (dash-dotted line). This distribution displays an overall
      morphology consistent with the driven turbulence
      simulations. \label{f:igTimePDFs} }
\end{figure}

Given that our initial conditions describe a region of fluid in a
uniform state with no velocity, no kernels exist at the initial
time. As turbulence is driven, the fluctuations in the hydrodynamic
state appear and grow in amplitude. This causes variations in ignition
times across the domain. Figure~\ref{f:igTimePDFs}
shows the distribution of mass as a function of ignition time for our
set of turbulence models. This form of the distribution is more
natural for discussing the evolution of ignition times as the system
parameters, such as the degree of compressibility in the driving,
change. The distributions were obtained as time averages over 3
integral time scales. In the case of purely solenoidal and 50 per cent
compressive driving, shown with dotted and dashed curves in
Figure~\ref{f:igTimePDFs}, respectively, the distributions are nearly
identical, and cannot be distinguished in the figure. These
distributions are nearly symmetric, with most mass residing near an
average ignition time of approximately \SI{1000}{\s}. They also have a
relatively small dispersion, with no material present (an amount less
than \num{e-3} of the peak mass) for ignition times shorter than about
\SI{1}{\s}. The mass distributions become progressively broader and
asymmetric as the degree of compressibility in the driving
increases. For example, the shortest ignition times observed in the
$C_0.75$ and $C_0.875$ models are approximately \SI{0.14}{\s} and
\SI{8e-3}{\s}. The distribution assumes a qualitatively different
shape in the case of the fully compressively driven model $C_1$. In
this case, the distribution is strongly asymmetric, with pronounced
tails extending down to approximately \SI{3e-7}{\s}.
\subsection{Spatial distribution of ignition times} \label{ss:spatialIgTimes}
\begin{figure*}
\setlength\figureheight{8cm}\ignorespaces 
\setlength\figurewidth{8cm}\ignorespaces
\begin{center}
    \begin{tabular}{ccc}
        \includegraphics{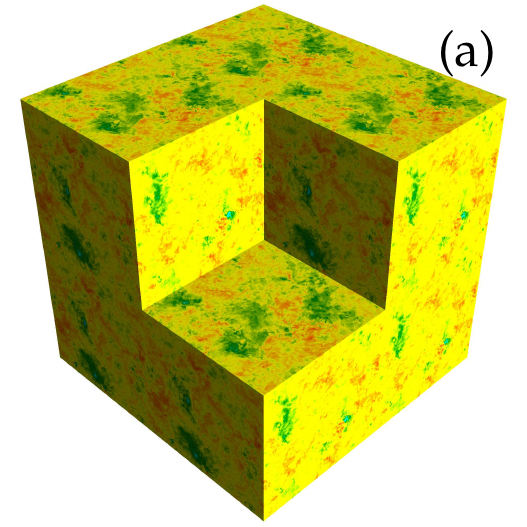}
        &
        \includegraphics{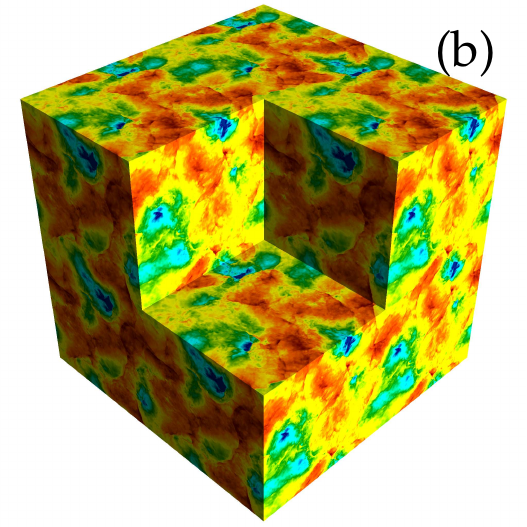}
        &
        \includegraphics{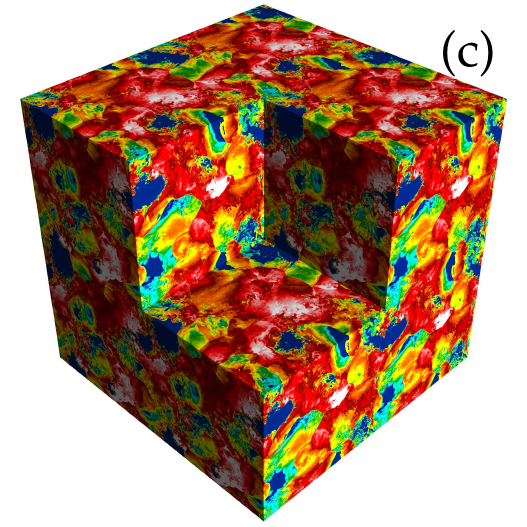}
    \end{tabular}
    \includegraphics{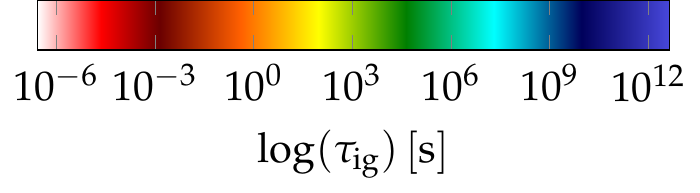}
    \caption
    { Pseudocolour maps showing ignition times for the interior and
      exterior surfaces for the $C_0$, $C_0.875$, and $C_1$
      models. Each volume shown here was created by replicating the
      original domain in the remaining octants of the Cartesian
      coordinate system whose origin lies at the corner of the
      original domain. Since our simulations use periodic boundary
      conditions, this results in continuous structures across the
      coordinate planes. The average colour in each model
      illustrates the extreme values of ignition times. Fluid
      structures also become increasingly distinct with increasing
      $\zeta_c$, as the prevalence of shocks
      increases. \label{f:volCutouts} }
\end{center}
\end{figure*}

Figure~\ref{f:volCutouts}
shows pseudocolour maps (in log scale) of ignition times for the $C_0$,
$C_0.75$ , and $C_1$ models, in the left, middle, and right panel,
respectively. When plotting the ignition times, we use the same scale,
which allows for making a qualitative comparison of spatial variations
in ignition times between models. In particular, the $C_0$ model is
characterized by the most uniform spatial distribution of ignition
times, with clearly-visible small-scale structure. This picture is
consistent with our analysis of the mass distributions as a function
of ignition times discussed in Section~\ref{s:massDistributions}. As
the degree of compressibility in the drive increases, however, the
spatial distribution of the ignition times becomes progressively more
non-uniform. In the case of the purely compressively-driven model,
$C_1$, there exist sharply-delineated regions of short and long
ignition times, resulting in an overall patchy appearance of the
distribution. This is due to the presence of shocks in strongly
compressively-driven models. In such cases, the ignition time is
relatively long in the region upstream of the shock, while it is the
shortest immediately behind the shock.

The spatial distributions of ignition times can be formally
characterized using spatial autocorrelation functions of ignition
times. We found that in the models in which driving was not strongly
compressive, the autocorrelation function provided evidence that
ignition times typically varied in structures spanning 5-6 mesh
cells. In contrast, the same analysis applied to the purely
compressively-driven model formally indicated the presence of
structures on scales below the mesh cell width. This result is
consistent with the patchy appearance of the distribution of ignition
times in this case, implying that the ignition times varied strongly
on small scales (separating individual `patches'). We point out that
integral scales comparable to or smaller than the mesh cell width
either point to the presence of under-resolved physics effects or a
failure of the analysis procedure. The presence of narrow regions
occupied by strong gradients of ignition times reflects on the
existence of shock fronts. In this case, the autocorrelation function
no longer provides information about the integral scale of the flow,
as the flow itself is no longer homogeneous, but composed of a
collection of shocked gas regions that cool and rarefy before they are
overrun by other shocks.
\subsection{Existence of detonation kernels}
As mentioned in Section~\ref{ss:analysisIgCond}, a detonation kernel
that will produce a self-sustained detonation must be characterized by
ignition times shorter than the local dynamical timescale, small
curvature, and must be able to preserve its integrity for long enough
for a detonation to fully develop. To this end, we applied the
clustering analysis (cf.\ Section~\ref{p:clustering}) to models $C_0$,
$C_{0.875}$, and $C_1$, using ignition time thresholds,
$\tau_{\mathrm{ign,pdk}}$, of \SI{e-5}{\s}, \SI{1}{\s}, and
\SI{100}{\s}, respectively. These ignition time thresholds were
selected in such a way that at least one prospective detonation kernel
was produced (in the case of the $C_0$ and $C_{0.875}$ models), and in
the case that many kernels were produced, they were spatially
well-separated (in the case of the $C_1$ model). Furthermore, we
disregarded any prospective kernels containing fewer than \num{e4}
mesh cells, as we deemed such small kernels as computationally
under-resolved.

\begin{figure*}
\adjustThreePanels\ignorespaces
\begin{center}
    \begin{tabular}{ccc} 
    \includegraphics{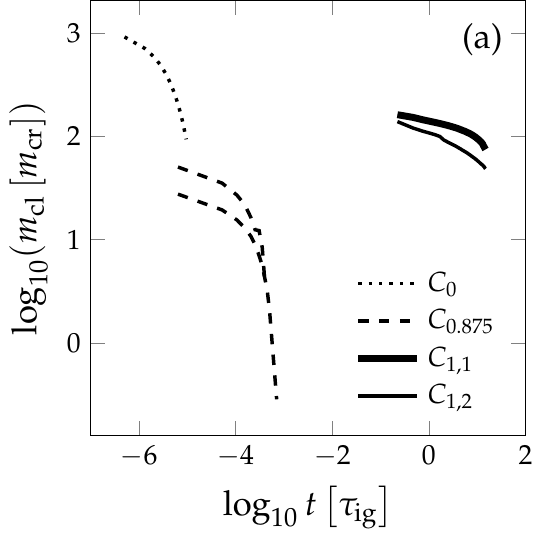}
    &
    \includegraphics{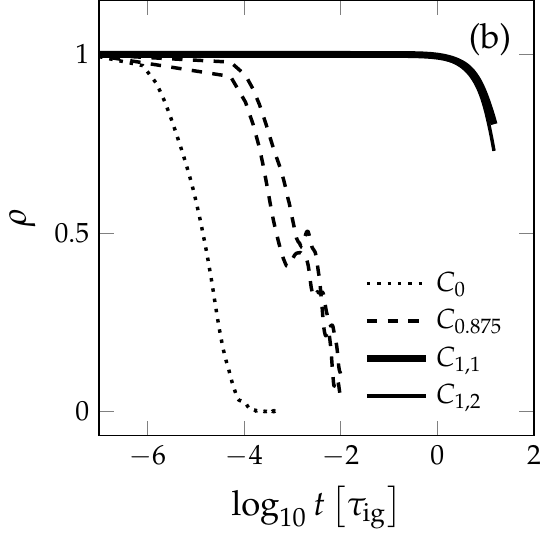}
    &
    \includegraphics{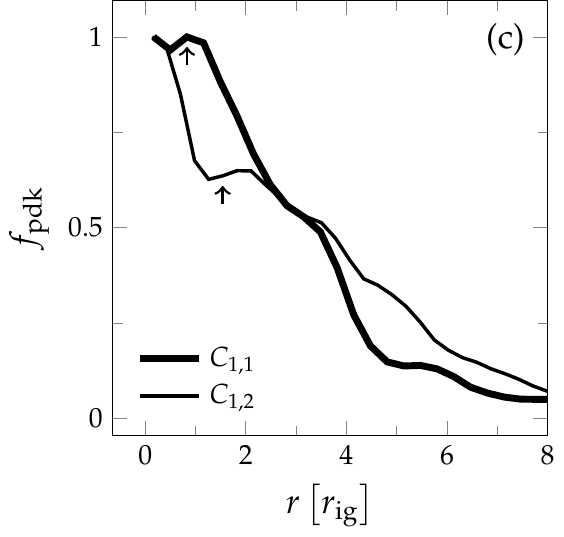}
    \end{tabular}
    \caption
    { Left panel: Eulerian analysis showing cluster masses and
      lifetimes for the models $C_0$ (dotted line), $C_0.875$ (dashed
      line), and $C_1$ (solid lines). For the latter two models, the
      largest two observed clusters are shown. The $C_0$ model
      developed only one cluster. The clusters in the $C_1$ model are
      likely ignition candidates, and are differentiated by the
      subscripts 1 and 2. No other models have clusters likely to
      ignite. Tracking of each cluster ceased when no unmixed original
      cluster material remained. Middle panel: Lagrangian cluster
      analysis. Autocorrelations of Lagrangian tracer particles
      originating in each cluster. Final times appear different from
      the Eulerian analysis because the particles retained their
      identity and could be tracked indefinitely. This analysis shows
      cluster dissolution due to mixing. Integral timescales for each
      cluster obtained by integrating the autocorrelation functions
      agree well with the cluster lifetimes found in the Eulerian
      analysis. Right panel: Filling factors for clusters from the
      $C_1$ model. The clusters shown are the same depicted in the
      other panels. The horizontal axis is normalized to the minimum
      ignition kernel radius (calculated using the average cluster
      density), for each cluster. Values of 1 indicate that a sphere
      of the given radius centred at the cluster's centre of mass is
      completely filled; values less than 1 indicate incomplete
      filling and departure from a spherical structure. The arrows
      indicate the minimum viable ignition kernel radius, calculated
      using the average density of the enclosed material, rather than
      the average cluster density. $C_{1,1}$ is completely filled at
      this radius, indicating it is a likely ignition
      candidate. $C_{1,2}$ is not, and thus is unlikely to produce a
      self-sustained detonation. \label{f:clusterEvolution} }
\end{center}        
\end{figure*}

Figure~\ref{f:clusterEvolution}(a)
shows the time evolution, in units of the corresponding threshold
ignition times, of the prospective detonation kernel mass. The kernel
mass is shown in log scale, and is normalized to the critical
mass. The critical mass is the mass of a spherical region with radius
equal to the critical radius, $r_{\mathrm{ig}}$, as given by equation
11 of \citet{dursi+06}, with $\beta=0.77$ and for a carbon
mass fraction of 0.5. In the calculation of the critical mass, we used
the average kernel mass density.

In the case of the purely solenoidally-driven model $C_0$, only one
cluster was found, and was destroyed very quickly after less than
\num{e-5} ignition times. We found two clusters in the $C_{0.875}$
model, which again were relatively quickly destroyed after less than
\num{e-3} ignition times. In contrast, the fully compressively-driven
model $C_1$ produced two clusters that survived essentially untouched
for more than 10 ignition times. It is worth pointing out that the
initial mass of the cluster, expressed in terms of the critical mass,
does not guarantee the survival of the cluster, as demonstrated in the
$C_0$ model, in which case the initial kernel mass was approximately
1000 times the critical mass.

We computed approximate prospective detonation kernel decorrelation
times by integrating the particle velocity autocorrelation functions
for each kernel
(cf.\ Section~\ref{p:longevity}). Figure~\ref{f:clusterEvolution}(b)
shows the time evolution of these velocity autocorrelation
functions. The typical thus-estimated decorrelation times were on the
order of \num{2e-5}, \num{3e-3}, and 13 ignition times for kernels
found in models $C_0$, $C_{0.875}$, and $C_1$, respectively. These
Lagrangian analysis results closely follow and confirm the results of
the Eulerian kernel longevity analysis.

Figure~\ref{f:clusterEvolution}(c) shows the filling factor of
prospective detonation kernels, $f_\mathrm{pdk}$, for the two kernels
found in the fully compressively-driven model. In the figure, the
radial dependence of the filling factor for $C_{1,1}$ and $C_{1,2}$
are shown with thick and thin solid lines, respectively. The radial
coordinate is scaled by the critical ignition radius,
$r_{\mathrm{ig}}$, as defined above, again using the average kernel
density. In the figure, the vertical arrows show the ignition radii
based on the actual density profile of the kernels. We note that these
radii differ from the critical kernel ignition radii estimated based
on the average kernel density, as described above. This is because the
central densities of the kernels are different from the average kernel
densities. Specifically, density in the central region of the
$C_{1,1}$ kernel is higher than the kernel's average density. In
contrast, the central region of the $C_{1,2}$ kernel is less dense
than this kernel's material on average. More importantly, the central
regions of the $C_{1,1}$ kernel up to twice the kernel ignition radius
are purely composed of material under ignition conditions. Although we
cannot exclude the possibility that the $C_{1,2}$ kernel will produce
a self-sustained detonation, the likelihood of such an outcome is much
greater in the case of the $C_{1,1}$ kernel.

It is interesting to note that the morphology of prospective
detonation kernels appears dependent on the compressibility of the
drive. Prospective kernels found in solenoidally-driven models are
characterized by a fractal dimension of about 2.62, while the fractal
dimension of prospective kernels found in purely compressively-driven
simulations is about 2.46. In general, these large fractal dimensions
are indicative of substantial structure of the kernels'
surfaces. Furthermore, compressively-driven models appear to produce
more organized flow. This is consistent with our qualitative
comparison of the flow morphology presented in
Section~\ref{ss:spatialIgTimes} (cf.\ Figure~\ref{f:volCutouts}).
\subsection{Application to white dwarf mergers}
To allow for direct comparison of results presented in this work to
simulations of white dwarf mergers, one must estimate the degree of
compressibility of turbulence present in the simulated boundary
layers. Only after such a connection is established will we be in a
position to identify additional factors of importance in Type Ia
explosions from merging white dwarfs.
\subsubsection{Relevance of the adopted model}
To provide an estimate of the compressibility typical of boundary
layers in white dwarf merger simulations, we extracted a cube of data
from the region in which a detonation formed in the He0609 model of
\citet{fenn+2016}. This region is depicted in Figure~10 of
\citet{fenn+2016}. The extracted data cube was \SI{256}{\km} per side,
which is a factor of 8 larger than the computational box used in the
present work. We used the merger model obtained about \SI{2}{\s} prior
to the formation of the detonation (for reference, the dynamical
timescale of the material in the extracted region was about
  \SI{1}{\s}).

Using the extracted data, we estimated the degree of compressibility
with help of the following procedure. In the first step, we decomposed
the velocity field into purely solenoidal and purely compressive
components \citep{balsara98}. This analysis indicated that the kinetic
energy stored in the compressive component amounted to
$\approx 37$ per cent of the total kinetic energy. Next, we
estimated the degree of compressibility in the driving in our merger
model by comparing the just-obtained amount of the kinetic energy
stored in the compressive mode to the corresponding amount found in
the database of turbulent simulations obtained in this work for
different degrees of compressibility, $\zeta_c$. We found that the
degree of compressibility with which the turbulence is driven in the
mergers is $\zeta_c \approx 0.97$. (A similar relation between the
degree of compressibility in the drive and the amount of kinetic
energy stored in the compressive component of the velocity field was
obtained by \citet{federrath+2010} in the context of modeling
turbulence in the interstellar medium.)

Although the drive is nearly purely-compressive in our He0609 model,
it is not clear whether such a drive will efficiently produce
detonation kernels. This is because of the strong dependence of
prospective detonation kernel properties on the drive compressibility
found in our simulations for $\zeta_c$ above 0.875. However, the
He0609 model \citep{fenn+2016} provides evidence for the formation of
detonations. Furthermore, the average pressure between a carbon-oxygen
mixture and pure helium, for the same density and temperature, differ
by less than about 17 per cent. This indicates that the
dynamics of the flow in the helium models should not be qualitatively
different from that in the carbon-oxygen models.

We compared the dependence of the mass distribution on the ignition
time in the He0609 model to our series of turbulent models. The
corresponding mass distribution in the He0609 model is shown with a
dot-dashed line in Figure~\ref{f:igTimePDFs}. In the figure, and for
clarity of presentation, the He0609 mass distribution has been
normalized to the maximum of the $C_1$ distribution, shown with a
thick solid line in Figure~\ref{f:igTimePDFs}. The corresponding
ignition times in the He0609 case were computed using the triple-alpha
burning timescale \citep{khokhlov+86}. Even though the He0609
distribution appears substantially different from that of the $C_1$
model, the difference in appearance is chiefly due to the difference
in the ignition times between helium and carbon-oxygen
mixtures. Indeed, closer analysis of the distribution shapes reveals
their qualitative similarity.

Table~\ref{t:distStats}
\ctable[
cap = Ignition time distribution statistics,
caption = {Skewness and kurtosis for the $C_{0.875}$, $C_1$, and merger boundary layer ignition time distributions} ,
label = t:distStats,
% mincapwidth = 7cm,
]{ccc}
{}
{\FL
model       & skewness  & kurtosis  \ML
$C_{0.875}$   & 0.46      & 3.38      \NN
$C_1$       & 0.89      & 3.52      \NN        
He0609      & 1.29      & 6.97      \LL  
}
provides information about the skewness and kurtosis for mass
distributions for a subset of relevant turbulence models and the
He0609 merger. It is interesting to note that the mass distribution in
the merger model is more strongly skewed and has heavier tails than
the mass distribution produced by the purely compressively-driven
$C_1$ model. This provides additional evidence for the similarity of
the flow found in the merger simulations to that obtained in
simulations of driven turbulence, justifying our approach. It also
offers evidence that successful detonations can be formed in systems
that are not driven purely compressively. In summary, we conclude that
the turbulence developing in white dwarf merger models has generic
properties reflected in dedicated driven turbulence models. The chief
difference between the two is the numerical mesh resolution.
\subsubsection{Key control parameters besides turbulence}
Thus, apart from the details of turbulence developing in the boundary
layer of the merging white dwarf binary, the thermonuclear ignition
time is the key remaining factor controlling the detonability of the
mixture. In turn, the ignition time chiefly depends on parameters of
the binary. Here the two important factors are the mass of the primary
component (that controls the depth of the potential well, and thus the
density and temperature of the accreted material), and the composition
of the companion star (that controls the composition of the accreted
material). The detonation kernels would exist provided that ignition
times found in the short ignition time tail of the mass distribution
are sufficiently short. This has important implications in regard to
the parameters of merging white dwarf binaries. The simulation-based
limit for detonability in binary white dwarf mergers--the total mass
of the system must exceed \SI{2.1}{\msun} \citep{dan+14}--was obtained
based on simulations that did not resolve turbulence. Thus, this limit
only considers the average background conditions rather than the
actual conditions that would be obtained in simulations resolving
turbulence. It is conceivable that if such a limit exists in nature,
it would correspond to a lower total system mass, and could be
established by performing merger simulations in which turbulence is
sufficiently resolved.
\subsubsection{Numerical effects}
The presence of the long-lived, large, and compact kernel $C_{1,1}$
offers evidence that self-sustained detonations can be produced in the
boundary layers of merging, carbon-oxygen white dwarfs, provided that
the primary is sufficiently massive. This argument is additionally
strengthened by the realization that our turbulence simulations
represent only a relatively small region of the boundary layer. One
may conservatively estimate, based on the typical linear span of the
large-scale vortices found in our merger models
\citep[ca.\ \SI{100}{\km},][]{fenn+2016}, compared to the size of our
computational box (\SI{32}{\km}), that there will potentially be
dozens of prospective detonation kernels inside a single vortex. In
addition, merger simulations indicate the presence of several large
vortices coexisting at any given time during the violent phase of the
merger. Therefore, it is expected that the boundary layer will be
densely populated with detonation kernels, making an explosion
essentially inevitable. However, the viable detonation kernel
$C_{1,1}$ has a radius of only about \SI{250}{\m}--significantly
smaller than the mesh resolution of any white dwarf merger simulations
performed to date. The fact that no detonations are observed in
carbon-oxygen boundary layers of systems with a total mass below
\SI{2.1}{\msun} can then be explained by mesh resolution inadequate to
resolve turbulence in simulations of merging white dwarfs binaries.
\subsection{Relation to the Khokhlov's semi-analytic detonation model}
\cite{khokhlov91b} considered a scenario in which a detonation is
created in the process of gradual strengthening of pre-existing flow
perturbations by plasma self-heating due to thermonuclear
burning. Because realistic multidimensional simulations were not
computationally feasible at the time of his study, Khokhlov adapted a
statistical formulation of \cite{meyer+71} to describe distribution of
plasma temperature fluctuations, or ignition times, on small
scales. He studied the evolution of ignition time distributions due to
thermonuclear self-heating, and concluded, in accordance with
\cite{meyer+71}, that a fluid parcel (prospective detonation kernel)
can produce a self-sustained detonation provided the energy release by
individual parts of the parcel are synchronized in time.

Our model offers an improvement upon Khokhlov's approach by
considering a realistic turbulent flow field, including the effects
due to the flow compressibility. We showed that in this case
distributions of flow perturbations, and thus the distributions of the
related ignition times, are no longer Gaussian, but become
progressively more asymmetric as the degree of compressibility in the
turbulent drive increases (cf.\ Figure~\ref{f:igTimePDFs} and
Table~\ref{t:distStats}). Also, compressibility qualitatively changes
the flow structure with ignition kernels isolated by relatively large
and smooth regions filled with expanding, shocked gas in the models
with strongly compressible drive (cf.\ Section
\ref{ss:spatialIgTimes}). Finally, time coherence of the ignition
centres, postulated by \cite{meyer+71} in order to produce strong
ignitions, is no longer needed as individual kernels considered in our
model are capable of producing self-sustained detonations.

Although we do not explicitly include self-heating due to
thermonuclear burning in our model, we demonstrated that such models
form a single family of solutions with models obtained at various
times differing only by the average background temperatures. This
implies that the mass distributions shown in Figure~\ref{f:igTimePDFs}
would, on average, undergo a horizontal shift toward shorter ignition
times as self-heating continues to increase the background
temperature. This is qualitatively consistent with the predicted by
Khokhlov evolution of ignition times \citep[see Section 3.4 and Figure
  6 in][]{khokhlov91b}.
\subsection{Application to explosions of Chandrasekhar-mass white dwarfs}
Here we consider the most popular SD scenario in which an explosion of
a Chandrasekhar-mass white dwarf is initiated in the process of a
deflagration. In the original model proposed by Nomoto and his
collaborators \citep{nomoto+76,nomoto+84}, the ignition occurs in the
centre of the white dwarf. Recent studies, however, indicate
that the large scale convection present in the core region before the
explosion may be responsible for advecting one or more nascent
deflagrating bubbles $\approx 100$ km away from the white
dwarf centre \citep{hoeflich+02,nonaka+12}. Due to buoyancy
effects, the following evolution strongly depends on how asymmetric
the initial conditions for the growth of the burning bubble are. If
the ignition occurs sufficiently close to the centre, the
burning region will grow and remain spherically-symmetric on average;
otherwise, buoyancy forces will induce a large scale acceleration of
the burning bubble \citep{niemeyer+96,malone+14}. In either case, the
flame is subjected to a range of instabilities, with the
Rayleigh-Taylor instability playing the dominant role in its
large-scale evolution \citep{khokhlov95,zhang+07}.
\subsubsection{Delayed detonation}
It is commonly accepted that as the flame reaches low-density
(${\approx}10^7$ g cm$^{-3}$) outer layers of the white
dwarf, it becomes susceptible to turbulence and fragments
\citep{hillebrandt+00,bell+04}. This particular phase of the flame
evolution has been considered as a possible source of a DDT originally
postulated by \cite{khokhlov91}. Unfortunately, the flow near the
flame region during that phase is subsonic \citep{roepke07}, and
well-resolved computer models failed to produce conditions conducive
to DDT \citep{zingale+05,aspden+10,aspden+11,woosley+11}. We cannot
exclude a possibility that a low Mach number approximation
\citep{bell+04} used in those dedicated, high-resolution simulations
prevented capturing important effects related to the flow
compressibility. But it is also possible that other physics components
omitted from consideration in the present study, such as plasma
magnetization and/or viscosity, act as additional reservoirs of energy
that under certain conditions could be tapped into to initiate
formation and support growth of the detonation kernels.
\subsubsection{Gravitationally Confined Detonation (GCD)} \label{ss:GCD}
GCD models follow the evolution of the deflagrating bubble after it is
ignited slightly off-centre \citep{plewa+04,seitenzahl+16},
in accordance with models of dense cores of massive white dwarfs prior
to the ignition \citep[][and references therein]{nonaka+12}. Although
shock-to-detonation transitions were observed in several model
realizations after the bubble material converged and collided in a
region opposite the breakout point \citep{plewa07,townsley+07},
turbulence and strong acoustic perturbations coexist in GCD models
also at earlier times. For example, a bow shock was observed forming
ahead of the deflagrating bubble prior to its breakout. This shock
sweeps through the surface layers, which very likely are turbulent due
to the combined effects of rotation, shell burning, and accretion.

Significant amount of turbulence is generated when the bubble material
moves over the surface, and mixes with and accelerates the outermost,
fuel-rich layers. Although it is difficult to expect a detonation
produced while the bubble ashes flow horizontally over the stellar
surface due to relatively low densities, a strongly mixed, turbulent
flow \citep[see][for a qualitative analysis of turbulence generated
  during this phase]{jordan+12} ultimately converges and collides
inside a region opposite the breakout point.

Apart from a possibility of a detonation triggered directly by a
shock, the plasma inside the convergence region is highly turbulent
with significant amount of strong, acoustic perturbations. Although
several 3D GCD models have been presented in the literature
\citep{roepke+07,jordan+08,seitenzahl+16}, meshes used in those
simulations did not have resolution necessary to describe turbulence
or resolve prospective detonation kernels.
\subsubsection{Pulsating Reverse Detonation (PRD)}
In the PRD scenario \citep[][and references therein]{bravo+09-prd2} a
deflagration is ignited in one or more regions in the core. In
contrast to GCD, however, the deflagrating material is contained
inside the star and the released energy is spent on inducing (radial)
pulsations of the white dwarf. After the initial expansion phase, the
star contracts and an accretion shock forms at some distance above the
stellar surface. This accretion shock eventually produces a converging
detonation wave.

Some comments on relation between GCD and PRD models are
due. Accretion shocks were observed in GCD models with relatively
energetic deflagration phases \citep[see, e.g. Y70YM25 and Y75YM50
  models in][]{plewa07}.\footnote{The energy produced by a typical
  deflagration in GCD \citep[$0.6-1.5\times 10^{50}$ erg,][]{plewa07}
  is by a factor 2--3 smaller than in PRD \citep[$2.5\times 10^{50}$
    erg,][]{bravo+06}.} Those models were characterized by relatively
stronger white dwarf expansion and slower surface flows, and therefore
were less conducive to prompt shock-to-detonation transitions. The
important difference is that the deflagrating bubbles in GCD retained
their structural integrity during their ascent from the core region to
the surface. It is not clear what physical mechanism could be
responsible for destruction of bubbles in PRD, and numerical effects
cannot be excluded.

Regardless qualitatively different outcome of the deflagration stage,
the PRD models feature a strongly perturbed surface layer that is
subjected to compression during the contraction phase. These are
conditions in which we predict detonations kernels will form.
\section{Summary}\label{s:summary}
We have performed a series of simulations of compressible, driven
turbulence under conditions representative of those found in the
boundary layer in white dwarf mergers with heavy primaries. In the
present study, these conditions correspond to a $0.8 +
\SI{1.2}{\solarmass}$ carbon-oxygen binary with an average density of
\SI{4e6}{\g\per\cubic\cm}, temperature of \SI{1e9}{\K}, and a 50/50
carbon-oxygen composition.

We used a spectral forcing method to drive turbulence in a cubic
domain, and studied the dependence of the likelihood of forming a
self-sustained detonation on the degree of compressibility in the
driving. We demonstrated the formal correctness of the procedure
adopted to model turbulence, and simulations displayed characteristics
(scaling of exponents of the velocity structure function, kinetic
energy spectra) consistent with those obtained in similar studies of
compressible turbulence.

We constructed mass distributions as a function of the mixture
ignition time. We identified and tracked the evolution of prospective
detonation kernels. We analysed these prospective kernels to quantify
their longevity, mass, and morphology in the context of
detonability. We also extracted data from the boundary layer region in
a simulation of merging white dwarfs (He+C/O), and drew comparisons
with our driven turbulence simulation results.

Our major findings and conclusions are summarized as follows:

\begin{enumerate}[i]

\item{Mass distributions as a function of ignition time indicate the
  presence of a heavy tail extending toward short ignition times,
  provided that the turbulence is driven compressively.}

\item{The analysis of the flow field of the ignition region extracted
  from the He0609 model of \citet{fenn+2016} shortly prior to helium
  ignition indicates that the turbulent drive does not need to be
  purely compressive in order to produce self-sustained
  detonations. The inferred degree of compressibility in the turbulent
  drive found in this model is about 97 per cent.}

\item{We demonstrated the overall similarity of the statistical
  properties of the mass distributions found in merger simulations and
  in our models. This justifies our approach to modeling the physics
  participating in the evolution of boundary layers in merging white
  dwarfs.}

\item{In our purely compressively-driven model, we found a substantial
  amount of material characterized by ignition times much shorter than
  the turbulent mixing timescale. Furthermore, we found this material
  organized in the form of centrally-condensed clusters. We identified
  these clusters as detonation kernels capable of producing
  self-sustained detonations.}

\item{The observed detonation kernel have a radius of about
  $\SI{250}{\m}$, which is notably smaller than the spatial resolution
  of any 3D white dwarf merger simulations performed to date. We
  conclude that, in order to correctly account for important physics
  participating in the process of merging binary white dwarfs,
  computer simulations must resolve flow structures on scales on the
  order of \SI{0.1}{\km}.}

\item{Our results indicate a high probability of detonations in the
  carbon-rich boundary layers of binary white dwarf systems with a
  total mass lower than previously advocated. In particular, the
  simulation-derived lower limit of the total white dwarf binary
  system mass of \SI{2.1}{\msun}, as suggested by \citet{dan+14}, may
  require revision. Adequately resolved simulations will likely allow
  for detonations in systems of lower total mass, thus broadening the
  population of binary white dwarf systems capable of producing Type
  Ia supernova explosions.}

\item{The proposed detonation mechanism also applies to single
  degenerates, and we identified possible sites harboring turbulence
  and strong acoustic perturbations in all popular explosion models of
  Chandrasekhar-mass white dwarfs. In the context of DDT mechanism in
  the delayed detonation model, we advocate the need for performing
  DNS studies of thermonuclear deflagrations at low densities using
  fully compressible hydrocodes.}

\end{enumerate}

The results of our study suggest a number of possible future research
directions. The average conditions adopted here correspond to
conditions characteristic of the boundary layer of a relatively
massive, $0.8 + \SI{1.2}{\solarmass}$, carbon-oxygen white dwarf
binary. It would be important to consider conditions corresponding to
systems with less massive carbon-oxygen primaries, and corroborate our
conclusions in regard to the minimal value of the total system mass
required for detonations. Simulations probing the parameter space of
average density, temperature, and composition should be conducted. As
we have recently shown \citep{fenn+2016}, self-heating from
nuclear energy release has an important effect on simulation
outcomes. Therefore, simulations of driven turbulence accounting for
the effects due to nuclear burning should be performed in order to
improve the conclusions drawn from the simplified model presented
here.
\section*{Acknowledgements}
We would like to acknowledge Tim Handy as the primary developer of the
\textsc{LAVAflow} analysis package, and we owe him special thanks for
valuable advice. We thank Sam Brenner for developing the
\textsc{LAVAflow} fractal analysis module. We would also like to thank
Phil Boehner for additional contributions to \textsc{LAVAflow}.

DF was supported by the DoD SMART Scholarship. TP was partially
supported by the NSF grant AST-1109113 and the DOE grant
DE-SC0008823. This research used resources of the National Energy
Research Scientific Computing Center, which is supported by the Office
of Science of the U.S.\ Department of Energy under Contract
No.\ DE-AC02-05CH11231, and the Air Force Research Laboratory. The
software used in this work was in part developed by the DOE Flash
Center at the University of Chicago. Data visualization was performed
in part using VisIt \citep{HPV:VisIt}. This research has made use of
NASA's Astrophysics Data System Bibliographic Services.

%%%%%%%%%%%%%%%%%%%%%%%%%%%%%%%%%%%%%%%%%%%%%%%%%%



\bibliographystyle{mnras}
\bibliography{manuscript}

%%%%%%%%%%%%%%%%%%%%%%%%%%%%%%%%%%%%%%%%%%%%%%%%%%

%%%%%%%%%%%%%%%%% APPENDICES %%%%%%%%%%%%%%%%%%%%%

%%%%%%%%%%%%%%%%%%%%%%%%%%%%%%%%%%%%%%%%%%%%%%%%%%

% Don't change these lines
\bsp	% typesetting comment
\label{lastpage}
\end{document}

%% file: packageIncludes.tex
\usepackage{amssymb}
\usepackage{fp}
\usepackage{xfrac}
\usepackage{amsmath}
\usepackage{hyperref}
\usepackage{booktabs}
\usepackage{cancel}
\usepackage{colortbl}
\usepackage{csquotes}
\usepackage{datatool}
\usepackage{helvet}
\usepackage{mathpazo}
\usepackage{listings}
\usepackage{multirow}
\usepackage{array}
\usepackage{float}

\usepackage{siunitx}

\usepackage{xparse}
\usepackage{etoolbox}
\usepackage{isotope}
\usepackage{ctable}
\usepackage{url}

%% file: figureMacros.tex
\newlength\figureheight 
\newlength\figurewidth 
\newcommand{\adjustThreePanels}{
    \setlength\figureheight{6cm}\ignorespaces 
    \setlength\figurewidth{6cm}\ignorespaces
    \setlength{\tabcolsep}{1pt}\ignorespaces
}

\newcommand{\adjustTwoPanels}{
    \setlength\figureheight{8cm}\ignorespaces 
    \setlength\figurewidth{8cm}\ignorespaces
    \setlength{\tabcolsep}{1pt}\ignorespaces
}

%% file: manuscript.bbl
\begin{thebibliography}{}
\makeatletter
\relax
\def\mn@urlcharsother{\let\do\@makeother \do\$\do\&\do\#\do\^\do\_\do\%\do\~}
\def\mn@doi{\begingroup\mn@urlcharsother \@ifnextchar [ {\mn@doi@}
  {\mn@doi@[]}}
\def\mn@doi@[#1]#2{\def\@tempa{#1}\ifx\@tempa\@empty \href
  {http://dx.doi.org/#2} {doi:#2}\else \href {http://dx.doi.org/#2} {#1}\fi
  \endgroup}
\def\mn@eprint#1#2{\mn@eprint@#1:#2::\@nil}
\def\mn@eprint@arXiv#1{\href {http://arxiv.org/abs/#1} {{\tt arXiv:#1}}}
\def\mn@eprint@dblp#1{\href {http://dblp.uni-trier.de/rec/bibtex/#1.xml}
  {dblp:#1}}
\def\mn@eprint@#1:#2:#3:#4\@nil{\def\@tempa {#1}\def\@tempb {#2}\def\@tempc
  {#3}\ifx \@tempc \@empty \let \@tempc \@tempb \let \@tempb \@tempa \fi \ifx
  \@tempb \@empty \def\@tempb {arXiv}\fi \@ifundefined
  {mn@eprint@\@tempb}{\@tempb:\@tempc}{\expandafter \expandafter \csname
  mn@eprint@\@tempb\endcsname \expandafter{\@tempc}}}

\bibitem[\protect\citeauthoryear{{Arnett}}{{Arnett}}{1996}]{arnett96}
{Arnett} W.~D.,  1996, Supernovae and nucleosynthesis. An investigation of the
  history of matter, from the Big Bang to the present.
Princeton University Press, Princeton

\bibitem[\protect\citeauthoryear{{Aspden}, {Bell}  \& {Woosley}}{{Aspden}
  et~al.}{2010}]{aspden+10}
{Aspden} A.~J.,  {Bell} J.~B.,   {Woosley} S.~E.,  2010, \mn@doi [\apj]
  {10.1088/0004-637X/710/2/1654}, \href
  {http://adsabs.harvard.edu/abs/2010ApJ...710.1654A} {710, 1654}

\bibitem[\protect\citeauthoryear{{Aspden}, {Bell}  \& {Woosley}}{{Aspden}
  et~al.}{2011}]{aspden+11}
{Aspden} A.~J.,  {Bell} J.~B.,   {Woosley} S.~E.,  2011, \mn@doi [\apj]
  {10.1088/0004-637X/730/2/144}, \href
  {http://adsabs.harvard.edu/abs/2011ApJ...730..144A} {730, 144}

\bibitem[\protect\citeauthoryear{{Balsara}}{{Balsara}}{1998}]{balsara98}
{Balsara} D.~S.,  1998, \mn@doi [\apjs] {10.1086/313093}, \href
  {http://adsabs.harvard.edu/abs/1998ApJS..116..133B} {116, 133}

\bibitem[\protect\citeauthoryear{{Bell}, {Day}, {Rendleman}, {Woosley}  \&
  {Zingale}}{{Bell} et~al.}{2004}]{bell+04}
{Bell} J.~B.,  {Day} M.~S.,  {Rendleman} C.~A.,  {Woosley} S.~E.,   {Zingale}
  M.~A.,  2004, \mn@doi [Journal of Computational Physics]
  {10.1016/j.jcp.2003.10.035}, \href
  {http://adsabs.harvard.edu/abs/2004JCoPh.195..677B} {195, 677}

\bibitem[\protect\citeauthoryear{{Blondin}, {Dessart}, {Hillier}  \&
  {Khokhlov}}{{Blondin} et~al.}{2013}]{blondin+13}
{Blondin} S.,  {Dessart} L.,  {Hillier} D.~J.,   {Khokhlov} A.~M.,  2013,
  \mn@doi [\mnras] {10.1093/mnras/sts484}, \href
  {http://adsabs.harvard.edu/abs/2013MNRAS.429.2127B} {429, 2127}

\bibitem[\protect\citeauthoryear{{Bravo} \& {Garc{\'{\i}}a-Senz}}{{Bravo} \&
  {Garc{\'{\i}}a-Senz}}{2006}]{bravo+06}
{Bravo} E.,  {Garc{\'{\i}}a-Senz} D.,  2006, \mn@doi [\apjl] {10.1086/504713},
  \href {http://adsabs.harvard.edu/abs/2006ApJ...642L.157B} {642, L157}

\bibitem[\protect\citeauthoryear{{Bravo}, {Garc{\'{\i}}a-Senz}, {Cabez{\'o}n}
  \& {Dom{\'{\i}}nguez}}{{Bravo} et~al.}{2009}]{bravo+09-prd2}
{Bravo} E.,  {Garc{\'{\i}}a-Senz} D.,  {Cabez{\'o}n} R.~M.,
  {Dom{\'{\i}}nguez} I.,  2009, \mn@doi [\apj] {10.1088/0004-637X/695/2/1257},
  \href {http://adsabs.harvard.edu/abs/2009ApJ...695.1257B} {695, 1257}

\bibitem[\protect\citeauthoryear{Childs et~al.,}{Childs
  et~al.}{2012}]{HPV:VisIt}
Childs H.,  et~al., 2012, in , {High Performance Visualization--Enabling
  Extreme-Scale Scientific Insight}.
pp 357--372

\bibitem[\protect\citeauthoryear{Colella \& Woodward}{Colella \&
  Woodward}{1984}]{colella+84}
Colella P.,  Woodward P.~R.,  1984, J. Comput. Phys., 54, 174

\bibitem[\protect\citeauthoryear{{Dan}, {Rosswog}, {Br{\"u}ggen}  \&
  {Podsiadlowski}}{{Dan} et~al.}{2014}]{dan+14}
{Dan} M.,  {Rosswog} S.,  {Br{\"u}ggen} M.,   {Podsiadlowski} P.,  2014,
  \mn@doi [\mnras] {10.1093/mnras/stt1766}, \href
  {http://adsabs.harvard.edu/abs/2014MNRAS.438...14D} {438, 14}

\bibitem[\protect\citeauthoryear{{Davidson}}{{Davidson}}{2004}]{davidson2004}
{Davidson} P.~A.,  2004, {Turbulence : an introduction for scientists and
  engineers}.
Oxford University Press, Oxford, UK

\bibitem[\protect\citeauthoryear{{Dunkley}, {Sharpe}  \& {Falle}}{{Dunkley}
  et~al.}{2013}]{dunkley+2013}
{Dunkley} S.~D.,  {Sharpe} G.~J.,   {Falle} S.~A.~E.~G.,  2013, \mn@doi
  [\mnras] {10.1093/mnras/stt422}, \href
  {http://adsabs.harvard.edu/abs/2013MNRAS.431.3429D} {431, 3429}

\bibitem[\protect\citeauthoryear{{Dursi} \& {Timmes}}{{Dursi} \&
  {Timmes}}{2006}]{dursi+06}
{Dursi} L.~J.,  {Timmes} F.~X.,  2006, \mn@doi [\apj] {10.1086/500638}, \href
  {http://adsabs.harvard.edu/abs/2006ApJ...641.1071D} {641, 1071}

\bibitem[\protect\citeauthoryear{{Eswaran} \& {Pope}}{{Eswaran} \&
  {Pope}}{1988}]{eswaran+88}
{Eswaran} V.,  {Pope} S.~B.,  1988, Computers and Fluids, \href
  {http://adsabs.harvard.edu/abs/1988CF.....16..257E} {16, 257}

\bibitem[\protect\citeauthoryear{Falconer}{Falconer}{2003}]{falconer+2003}
Falconer K.~J.,  2003, Fractal geometry : mathematical foundations and
  applications.
J. Wiley \& Sons, Chichester, New York

\bibitem[\protect\citeauthoryear{{Federrath}, {Roman-Duval}, {Klessen},
  {Schmidt}  \& {Mac Low}}{{Federrath} et~al.}{2010}]{federrath+2010}
{Federrath} C.,  {Roman-Duval} J.,  {Klessen} R.~S.,  {Schmidt} W.,   {Mac Low}
  M.-M.,  2010, \mn@doi [A&A] {10.1051/0004-6361/200912437}, 512, A81

\bibitem[\protect\citeauthoryear{{Fenn}, {Plewa}  \& {Gawryszczak}}{{Fenn}
  et~al.}{2016}]{fenn+2016}
{Fenn} D.,  {Plewa} T.,   {Gawryszczak} A.,  2016, \mn@doi [\mnras]
  {10.1093/mnras/stw1831}, \href
  {http://adsabs.harvard.edu/abs/2016MNRAS.462.2486F} {462, 2486}

\bibitem[\protect\citeauthoryear{Fisher et~al.,}{Fisher
  et~al.}{2008}]{fisher+2008}
Fisher R.~T.,  et~al., 2008, IBM Journal of Research and Development, 52, 127

\bibitem[\protect\citeauthoryear{{Fowler} \& {Hoyle}}{{Fowler} \&
  {Hoyle}}{1964}]{fowler+64}
{Fowler} W.~A.,  {Hoyle} F.,  1964, \mn@doi [\apjs] {10.1086/190103}, \href
  {http://adsabs.harvard.edu/abs/1964ApJS....9..201F} {9, 201}

\bibitem[\protect\citeauthoryear{{Fryxell} et~al.,}{{Fryxell}
  et~al.}{2000}]{Fryxell+00}
{Fryxell} B.,  et~al., 2000, \apjs, 131, 273

\bibitem[\protect\citeauthoryear{{Gamezo}, {Khokhlov}  \& {Oran}}{{Gamezo}
  et~al.}{2005}]{gamezo+05}
{Gamezo} V.~N.,  {Khokhlov} A.~M.,   {Oran} E.~S.,  2005, \mn@doi [\apj]
  {10.1086/428767}, \href {http://adsabs.harvard.edu/abs/2005ApJ...623..337G}
  {623, 337}

\bibitem[\protect\citeauthoryear{{Guillochon}, {Dan}, {Ramirez-Ruiz}  \&
  {Rosswog}}{{Guillochon} et~al.}{2010}]{guillochon+10}
{Guillochon} J.,  {Dan} M.,  {Ramirez-Ruiz} E.,   {Rosswog} S.,  2010, \mn@doi
  [\apjl] {10.1088/2041-8205/709/1/L64}, \href
  {http://adsabs.harvard.edu/abs/2010ApJ...709L..64G} {709, L64}

\bibitem[\protect\citeauthoryear{{Hillebrandt} \& {Niemeyer}}{{Hillebrandt} \&
  {Niemeyer}}{2000}]{hillebrandt+00}
{Hillebrandt} W.,  {Niemeyer} J.~C.,  2000, \mn@doi [\araa]
  {10.1146/annurev.astro.38.1.191}, \href
  {http://adsabs.harvard.edu/abs/2000ARA%26A..38..191H} {38, 191}

\bibitem[\protect\citeauthoryear{{H{\"o}flich} \& {Stein}}{{H{\"o}flich} \&
  {Stein}}{2002}]{hoeflich+02}
{H{\"o}flich} P.,  {Stein} J.,  2002, \mn@doi [\apj] {10.1086/338981}, \href
  {http://adsabs.harvard.edu/abs/2002ApJ...568..779H} {568, 779}

\bibitem[\protect\citeauthoryear{{Hoyle} \& {Fowler}}{{Hoyle} \&
  {Fowler}}{1960}]{Hoyle_Fowler}
{Hoyle} F.,  {Fowler} W.~A.,  1960, \mn@doi [\apj] {10.1086/146963}, \href
  {http://adsabs.harvard.edu/abs/1960ApJ...132..565H} {132, 565}

\bibitem[\protect\citeauthoryear{{Jordan}, {Fisher}, {Townsley}, {Calder},
  {Graziani}, {Asida}, {Lamb}  \& {Truran}}{{Jordan} et~al.}{2008}]{jordan+08}
{Jordan} IV G.~C.,  {Fisher} R.~T.,  {Townsley} D.~M.,  {Calder} A.~C.,
  {Graziani} C.,  {Asida} S.,  {Lamb} D.~Q.,   {Truran} J.~W.,  2008, \mn@doi
  [\apj] {10.1086/588269}, \href
  {http://adsabs.harvard.edu/abs/2008ApJ...681.1448J} {681, 1448}

\bibitem[\protect\citeauthoryear{{Jordan} IV et~al.,}{{Jordan}
  et~al.}{2012}]{jordan+12}
{Jordan} IV G.~C.,  et~al., 2012, \mn@doi [\apj] {10.1088/0004-637X/759/1/53},
  \href {http://adsabs.harvard.edu/abs/2012ApJ...759...53J} {759, 53}

\bibitem[\protect\citeauthoryear{{Kashyap}, {Fisher}, {Garc{\'{\i}}a-Berro},
  {Aznar-Sigu{\'a}n}, {Ji}  \& {Lor{\'e}n-Aguilar}}{{Kashyap}
  et~al.}{2015}]{kashyap+2015}
{Kashyap} R.,  {Fisher} R.,  {Garc{\'{\i}}a-Berro} E.,  {Aznar-Sigu{\'a}n} G.,
  {Ji} S.,   {Lor{\'e}n-Aguilar} P.,  2015, \mn@doi [\apjl]
  {10.1088/2041-8205/800/1/L7}, \href
  {http://adsabs.harvard.edu/abs/2015ApJ...800L...7K} {800, L7}

\bibitem[\protect\citeauthoryear{{Khokhlov}}{{Khokhlov}}{1991a}]{khokhlov91}
{Khokhlov} A.~M.,  1991a, \aap, \href
  {http://adsabs.harvard.edu/abs/1991A%26A...245..114K} {245, 114}

\bibitem[\protect\citeauthoryear{{Khokhlov}}{{Khokhlov}}{1991b}]{khokhlov91b}
{Khokhlov} A.~M.,  1991b, \aap, \href
  {http://adsabs.harvard.edu/abs/1991A%26A...246..383K} {246, 383}

\bibitem[\protect\citeauthoryear{{Khokhlov}}{{Khokhlov}}{1995}]{khokhlov95}
{Khokhlov} A.~M.,  1995, \mn@doi [\apj] {10.1086/176091}, \href
  {http://adsabs.harvard.edu/abs/1995ApJ...449..695K} {449, 695}

\bibitem[\protect\citeauthoryear{{Khokhlov} \& {Ergma}}{{Khokhlov} \&
  {Ergma}}{1986}]{khokhlov+86}
{Khokhlov} A.~M.,  {Ergma} E.~V.,  1986, Soviet Astronomy Letters, \href
  {http://adsabs.harvard.edu/abs/1986SvAL...12..152K} {12, 152}

\bibitem[\protect\citeauthoryear{{Khokhlov}, {Oran}  \& {Wheeler}}{{Khokhlov}
  et~al.}{1997}]{khokhlov+97}
{Khokhlov} A.~M.,  {Oran} E.~S.,   {Wheeler} J.~C.,  1997, \apj, \href
  {http://adsabs.harvard.edu/abs/1997ApJ...478..678K} {478, 678}

\bibitem[\protect\citeauthoryear{Levoy}{Levoy}{1981}]{levoy81}
Levoy M.,  1981, in SIGGRAPH '81. pp 6--12

\bibitem[\protect\citeauthoryear{{Malone}, {Nonaka}, {Woosley}, {Almgren},
  {Bell}, {Dong}  \& {Zingale}}{{Malone} et~al.}{2014}]{malone+14}
{Malone} C.~M.,  {Nonaka} A.,  {Woosley} S.~E.,  {Almgren} A.~S.,  {Bell}
  J.~B.,  {Dong} S.,   {Zingale} M.,  2014, \mn@doi [\apj]
  {10.1088/0004-637X/782/1/11}, \href
  {http://adsabs.harvard.edu/abs/2014ApJ...782...11M} {782, 11}

\bibitem[\protect\citeauthoryear{{Meyer} \& {Oppenheim}}{{Meyer} \&
  {Oppenheim}}{1971}]{meyer+71}
{Meyer} J.~W.,  {Oppenheim} A.~K.,  1971, \mn@doi [Combust. Flame]
  {http://dx.doi.org/10.1016/S0010-2180(71)80139-6}, 17, 65

\bibitem[\protect\citeauthoryear{{Mochkovitch} \& {Livio}}{{Mochkovitch} \&
  {Livio}}{1989}]{mochkovitch+89}
{Mochkovitch} R.,  {Livio} M.,  1989, \aap, \href
  {http://adsabs.harvard.edu/abs/1989A%26A...209..111M} {209, 111}

\bibitem[\protect\citeauthoryear{{Moll}, {Raskin}, {Kasen}  \&
  {Woosley}}{{Moll} et~al.}{2014}]{moll+2014}
{Moll} R.,  {Raskin} C.,  {Kasen} D.,   {Woosley} S.~E.,  2014, \mn@doi [\apj]
  {10.1088/0004-637X/785/2/105}, \href
  {http://adsabs.harvard.edu/abs/2014ApJ...785..105M} {785, 105}

\bibitem[\protect\citeauthoryear{{Niemeyer}, {Hillebrandt}  \&
  {Woosley}}{{Niemeyer} et~al.}{1996}]{niemeyer+96}
{Niemeyer} J.~C.,  {Hillebrandt} W.,   {Woosley} S.~E.,  1996, \mn@doi [\apj]
  {10.1086/178017}, \href {http://adsabs.harvard.edu/abs/1996ApJ...471..903N}
  {471, 903}

\bibitem[\protect\citeauthoryear{{Nomoto}, {Sugimoto}  \& {Neo}}{{Nomoto}
  et~al.}{1976}]{nomoto+76}
{Nomoto} K.,  {Sugimoto} D.,   {Neo} S.,  1976, \mn@doi [\apss]
  {10.1007/BF00648354}, \href
  {http://adsabs.harvard.edu/abs/1976Ap%26SS..39L..37N} {39, L37}

\bibitem[\protect\citeauthoryear{{Nomoto}, {Thielemann}  \& {Yokoi}}{{Nomoto}
  et~al.}{1984}]{nomoto+84}
{Nomoto} K.,  {Thielemann} F.-K.,   {Yokoi} K.,  1984, \mn@doi [\apj]
  {10.1086/162639}, \href {http://adsabs.harvard.edu/abs/1984ApJ...286..644N}
  {286, 644}

\bibitem[\protect\citeauthoryear{{Nonaka}, {Aspden}, {Zingale}, {Almgren},
  {Bell}  \& {Woosley}}{{Nonaka} et~al.}{2012}]{nonaka+12}
{Nonaka} A.,  {Aspden} A.~J.,  {Zingale} M.,  {Almgren} A.~S.,  {Bell} J.~B.,
  {Woosley} S.~E.,  2012, \mn@doi [\apj] {10.1088/0004-637X/745/1/73}, \href
  {http://adsabs.harvard.edu/abs/2012ApJ...745...73N} {745, 73}

\bibitem[\protect\citeauthoryear{Oran \& Gamezo}{Oran \&
  Gamezo}{2007}]{oran+07}
Oran E.~S.,  Gamezo V.~N.,  2007, \mn@doi [Combustion and Flame]
  {http://dx.doi.org/10.1016/j.combustflame.2006.07.010}, 148, 4

\bibitem[\protect\citeauthoryear{{Pakmor}, {Kromer}, {R{\"o}pke}, {Sim},
  {Ruiter}  \& {Hillebrandt}}{{Pakmor} et~al.}{2010}]{pakmor+10}
{Pakmor} R.,  {Kromer} M.,  {R{\"o}pke} F.~K.,  {Sim} S.~A.,  {Ruiter} A.~J.,
  {Hillebrandt} W.,  2010, \mn@doi [\nat] {10.1038/nature08642}, \href
  {http://adsabs.harvard.edu/abs/2010Natur.463...61P} {463, 61}

\bibitem[\protect\citeauthoryear{{Plewa}}{{Plewa}}{2007}]{plewa07}
{Plewa} T.,  2007, \mn@doi [\apj] {10.1086/511412}, \href
  {http://adsabs.harvard.edu/abs/2007ApJ...657..942P} {657, 942}

\bibitem[\protect\citeauthoryear{{Plewa}, {Calder}  \& {Lamb}}{{Plewa}
  et~al.}{2004}]{plewa+04}
{Plewa} T.,  {Calder} A.~C.,   {Lamb} D.~Q.,  2004, \mn@doi [\apjl]
  {10.1086/424036}, \href {http://adsabs.harvard.edu/abs/2004ApJ...612L..37P}
  {612, L37}

\bibitem[\protect\citeauthoryear{{Poludnenko}, {Gardiner}  \&
  {Oran}}{{Poludnenko} et~al.}{2011}]{poludnenko+11}
{Poludnenko} A.~Y.,  {Gardiner} T.~A.,   {Oran} E.~S.,  2011, \mn@doi [Physical
  Review Letters] {10.1103/PhysRevLett.107.054501}, \href
  {http://adsabs.harvard.edu/abs/2011PhRvL.107e4501P} {107, 054501}

\bibitem[\protect\citeauthoryear{{Raskin}, {Kasen}, {Moll}, {Schwab}  \&
  {Woosley}}{{Raskin} et~al.}{2014}]{raskin+2014}
{Raskin} C.,  {Kasen} D.,  {Moll} R.,  {Schwab} J.,   {Woosley} S.,  2014,
  \mn@doi [\apj] {10.1088/0004-637X/788/1/75}, \href
  {http://adsabs.harvard.edu/abs/2014ApJ...788...75R} {788, 75}

\bibitem[\protect\citeauthoryear{{R{\"o}pke}}{{R{\"o}pke}}{2007}]{roepke07}
{R{\"o}pke} F.~K.,  2007, \mn@doi [\apj] {10.1086/520830}, \href
  {http://adsabs.harvard.edu/abs/2007ApJ...668.1103R} {668, 1103}

\bibitem[\protect\citeauthoryear{{R{\"o}pke}, {Woosley}  \&
  {Hillebrandt}}{{R{\"o}pke} et~al.}{2007}]{roepke+07}
{R{\"o}pke} F.~K.,  {Woosley} S.~E.,   {Hillebrandt} W.,  2007, \mn@doi [\apj]
  {10.1086/512769}, \href {http://adsabs.harvard.edu/abs/2007ApJ...660.1344R}
  {660, 1344}

\bibitem[\protect\citeauthoryear{{Seitenzahl}, {Meakin}, {Townsley}, {Lamb}  \&
  {Truran}}{{Seitenzahl} et~al.}{2009}]{seitenzahl+2009}
{Seitenzahl} I.~R.,  {Meakin} C.~A.,  {Townsley} D.~M.,  {Lamb} D.~Q.,
  {Truran} J.~W.,  2009, \mn@doi [\apj] {10.1088/0004-637X/696/1/515}, \href
  {http://adsabs.harvard.edu/abs/2009ApJ...696..515S} {696, 515}

\bibitem[\protect\citeauthoryear{{Seitenzahl} et~al.,}{{Seitenzahl}
  et~al.}{2016}]{seitenzahl+16}
{Seitenzahl} I.~R.,  et~al., 2016, \mn@doi [\aap]
  {10.1051/0004-6361/201527251}, \href
  {http://adsabs.harvard.edu/abs/2016A%26A...592A..57S} {592, A57}

\bibitem[\protect\citeauthoryear{{Sharpe}}{{Sharpe}}{2001}]{sharpe+01}
{Sharpe} G.~J.,  2001, \mn@doi [\mnras] {10.1046/j.1365-8711.2001.04119.x},
  \href {http://adsabs.harvard.edu/abs/2001MNRAS.322..614S} {322, 614}

\bibitem[\protect\citeauthoryear{{She} \& {Leveque}}{{She} \&
  {Leveque}}{1994}]{she+94}
{She} Z.-S.,  {Leveque} E.,  1994, \mn@doi [Physical Review Letters]
  {10.1103/PhysRevLett.72.336}, \href
  {http://adsabs.harvard.edu/abs/1994PhRvL..72..336S} {72, 336}

\bibitem[\protect\citeauthoryear{{Tennekes} \& {Lumley}}{{Tennekes} \&
  {Lumley}}{1972}]{tennekes+72}
{Tennekes} H.,  {Lumley} J.~L.,  1972, {First Course in Turbulence}.
MIT Press, Cambridge

\bibitem[\protect\citeauthoryear{{Timmes} \& {Swesty}}{{Timmes} \&
  {Swesty}}{2000}]{timmes+2000}
{Timmes} F.~X.,  {Swesty} F.~D.,  2000, \mn@doi [\apjs] {10.1086/313304}, \href
  {http://adsabs.harvard.edu/abs/2000ApJS..126..501T} {126, 501}

\bibitem[\protect\citeauthoryear{{Townsley}, {Calder}, {Asida}, {Seitenzahl},
  {Peng}, {Vladimirova}, {Lamb}  \& {Truran}}{{Townsley}
  et~al.}{2007}]{townsley+07}
{Townsley} D.~M.,  {Calder} A.~C.,  {Asida} S.~M.,  {Seitenzahl} I.~R.,  {Peng}
  F.,  {Vladimirova} N.,  {Lamb} D.~Q.,   {Truran} J.~W.,  2007, \mn@doi [\apj]
  {10.1086/521013}, \href {http://adsabs.harvard.edu/abs/2007ApJ...668.1118T}
  {668, 1118}

\bibitem[\protect\citeauthoryear{{Woosley}, {Kerstein}  \& {Aspden}}{{Woosley}
  et~al.}{2011a}]{woosley+2011}
{Woosley} S.~E.,  {Kerstein} A.~R.,   {Aspden} A.~J.,  2011a, \mn@doi [\apj]
  {10.1088/0004-637X/734/1/37}, \href
  {http://adsabs.harvard.edu/abs/2011ApJ...734...37W} {734, 37}

\bibitem[\protect\citeauthoryear{{Woosley}, {Kerstein}  \& {Aspden}}{{Woosley}
  et~al.}{2011b}]{woosley+11}
{Woosley} S.~E.,  {Kerstein} A.~R.,   {Aspden} A.~J.,  2011b, \mn@doi [\apj]
  {10.1088/0004-637X/734/1/37}, \href
  {http://adsabs.harvard.edu/abs/2011ApJ...734...37W} {734, 37}

\bibitem[\protect\citeauthoryear{{Yungelson} \& {Kuranov}}{{Yungelson} \&
  {Kuranov}}{2017}]{yungelson+17}
{Yungelson} L.~R.,  {Kuranov} A.~G.,  2017, \mn@doi [\mnras]
  {10.1093/mnras/stw2432}, \href
  {http://adsabs.harvard.edu/abs/2017MNRAS.464.1607Y} {464, 1607}

\bibitem[\protect\citeauthoryear{{Zel'dovich}, {Librovich}, {Makhviladze}  \&
  {Sivashinskil}}{{Zel'dovich} et~al.}{1970}]{zeldovich+70}
{Zel'dovich} Y.~B.,  {Librovich} V.~B.,  {Makhviladze} G.~M.,   {Sivashinskil}
  G.~I.,  1970, \mn@doi [Journal of Applied Mechanics and Technical Physics]
  {10.1007/BF00908106}, \href
  {http://adsabs.harvard.edu/abs/1970JAMTP..11..264Z} {11, 264}

\bibitem[\protect\citeauthoryear{{Zhang}, {Messer}, {Khokhlov}  \&
  {Plewa}}{{Zhang} et~al.}{2007}]{zhang+07}
{Zhang} J.,  {Messer} O.~E.~B.,  {Khokhlov} A.~M.,   {Plewa} T.,  2007, \mn@doi
  [\apj] {10.1086/510145}, \href
  {http://adsabs.harvard.edu/abs/2007ApJ...656..347Z} {656, 347}

\bibitem[\protect\citeauthoryear{{Zingale}, {Woosley}, {Rendleman}, {Day}  \&
  {Bell}}{{Zingale} et~al.}{2005}]{zingale+05}
{Zingale} M.,  {Woosley} S.~E.,  {Rendleman} C.~A.,  {Day} M.~S.,   {Bell}
  J.~B.,  2005, \mn@doi [\apj] {10.1086/433164}, \href
  {http://adsabs.harvard.edu/abs/2005ApJ...632.1021Z} {632, 1021}

\makeatother
\end{thebibliography}
